\documentclass{llncs} %

\usepackage{float}
\usepackage{graphicx}
\usepackage{wrapfig}

\usepackage[labelfont=bf]{caption}
\usepackage[colorlinks=true,linkcolor=blue,citecolor=blue]{hyperref}

\usepackage{color,soul}
\usepackage{amsmath} 
\usepackage{mathtools} 
\usepackage{amsfonts} 
\usepackage{amssymb}
 
\usepackage{array} 
\usepackage{booktabs}
\usepackage{rotating}
\usepackage{multirow}
\usepackage{multicol}
\usepackage{lscape}
\usepackage{subfig}
\usepackage{times}

\usepackage[export]{adjustbox}

\usepackage[ruled,vlined]{algorithm2e}

\usepackage{xargs}                      

\usepackage[colorinlistoftodos,prependcaption,textsize=tiny]{todonotes}

\usepackage{tikz,xcolor,hyperref}

\definecolor{lime}{HTML}{A6CE39}
\DeclareRobustCommand{\orcidicon}{
	\begin{tikzpicture}
	\draw[lime, fill=lime] (0,0) 
	circle [radius=0.16] 
	node[white] {{\fontfamily{qag}\selectfont \tiny ID}};
	\draw[white, fill=white] (-0.0625,0.095) 
	circle [radius=0.007];
	\end{tikzpicture}
	\hspace{-2mm}
}

\foreach \x in {A, ..., Z}{\expandafter\xdef\csname orcid\x\endcsname{\noexpand\href{https://orcid.org/\csname orcidauthor\x\endcsname}
			{\noexpand\orcidicon}}
}
			
\definecolor{darkgreen}{rgb}{0.53, 0.66, 0.42}

\begin{document}

\title{Non-isomorphic Inter-modality Graph Alignment and Synthesis for Holistic Brain Mapping}

\titlerunning{Short Title}  

\author{Islem Mhiri\orcidC{}\index{Mhiri, Islem}\inst{1,2} \and Ahmed Nebli\orcidB{}\index{Nebli, Ahmed}\inst{1,2} \and Mohamed Ali Mahjoub\orcidD{} \index{Mahjoub, Mohamed Ali}\inst{1} \and \\ Islem Rekik\orcidA{} \index{Rekik, Islem}\inst{2}\thanks{ {corresponding author: \url{irekik@itu.edu.tr}.}} }

\authorrunning{I Mhiri et al.}

\institute{$^{1}$ Universit\'e de Sousse, Ecole Nationale d'Ing\'enieurs de Sousse, LATIS- Laboratory of Advanced Technology and Intelligent Systems, 4023, Sousse, Tunisie;\\
$^{2}$ BASIRA Lab, Faculty of Computer and Informatics Engineering, Istanbul Technical University, Istanbul, Turkey (\url{http://basira-lab.com/})}

\maketitle              

\begin{abstract}
Brain graph synthesis marked a new era for predicting a target brain graph from a source one without incurring the high acquisition cost and processing time of neuroimaging data. However, works on recovering a brain graph in one modality (e.g., functional brain imaging) from a brain graph in another (e.g., structural brain imaging) remain largely scarce. Besides, existing \emph{multimodal} graph synthesis frameworks have several limitations. \emph{First}, they mainly focus on generating graphs from the same domain (intra-modality), overlooking the rich multimodal representations of brain connectivity (inter-modality). \emph{Second}, they can only handle \emph{isomorphic} graph generation tasks, limiting their generalizability to synthesizing target graphs with a different node size and topological structure from those of the source one. More importantly, both target and source domains might have different distributions, which causes a domain fracture between them (i.e., distribution misalignment). To address such challenges, we propose an inter-modality aligner of non-isomorphic graphs (IMANGraphNet) framework to infer a target graph modality based on a given modality. Our three core contributions lie in (i) predicting a target graph (e.g., functional) from a source graph (e.g., morphological) based on a novel graph generative adversarial network (gGAN); (ii) using non-isomorphic graphs for both source and target domains with a different number of nodes, edges and structure; and (iii) enforcing the source distribution to match that of the ground truth graphs using a graph aligner to relax the loss function to optimize. Furthermore, to handle the unstable behavior of gGAN, we design a new Ground Truth-Preserving (GT-P) loss function to guide the non-isomorphic generator in learning the topological structure of ground truth brain graphs more effectively. Our comprehensive experiments on predicting target functional brain graphs from source morphological graphs demonstrate the outperformance of IMANGraphNet in comparison with its variants. IMANGraphNet presents the first framework for brain graph synthesis based on aligned non-isomorphic inter-modality brain graphs which handles variations in graph size, distribution and structure. This can be further leveraged for integrative and holistic brain mapping as well as developing multimodal neurological disorder diagnosis frameworks. Our Python IMANGraphNet code is available at \url{https://github.com/basiralab/IMANGraphNet}\footnote{\textbf{Paper YouTube video:} \url{https://www.youtube.com/watch?v=kzS-PkOt5_4&ab_channel=BASIRALab}}.
\end{abstract}


\section{Introduction}
Multimodal brain imaging spinned several medical image analysis applications thanks to the rich multimodal information it provides \cite{shen2013,zhang2019}. Multiple data sources such as magnetic resonance imaging (MRI), computed tomography (CT), and positron emission tomography (PET) offer the possibility of learning more holistic and informative data representations. However, such modalities introduce challenges including high acquisition cost and processing time across different clinical facilities.

Following the exponential growth of deep learning applications using MRI data, recently, such end-to-end frameworks have been investigated for multimodal MR image synthesis \cite{yu2020,zhou2020}. These methods either synthesize one modality from another (i.e., cross-modality) or map both modalities to a commonly shared domain.
Specifically, generative adversarial networks (GANs) have held great promise in predicting medical images of different brain image modalities from a given modality \cite{liu2020,dai2020,yang2020}. For instance, \cite{liu2020} suggested a joint neuroimage synthesis and representation learning (JSRL) framework with transfer learning for subjective cognitive decline conversion prediction where they imputed missing PET images using MRI scans. In addition, \cite{dai2020} proposed a unified GAN to train only a single generator and a single discriminator to learn the mappings among images of four different modalities. Furthermore, \cite{yang2020} translated a T1-weighted magnetic resonance imaging (MRI) to T2-weighted MRI using GAN. Although significant clinical representations were obtained from the latter studies, more substantial challenges still exist \cite{zhang2019}. As the reader may recognize, the brain connectome is a complex non-linear structure, which makes it difficult to be captured by linear models \cite{bassett2017,van2019}. Besides, many methods do not make good use of or even fail to treat non-euclidian structured data (i.e., geometric data) types such as graphs and manifolds \cite{bronstein2017}. Hence a deep learning model that preserves the topology of graph-based data representations for the target learning task presents a pertinent research direction to explore. 

\begin{figure}[ht!]
\centering
\includegraphics[width=11cm]{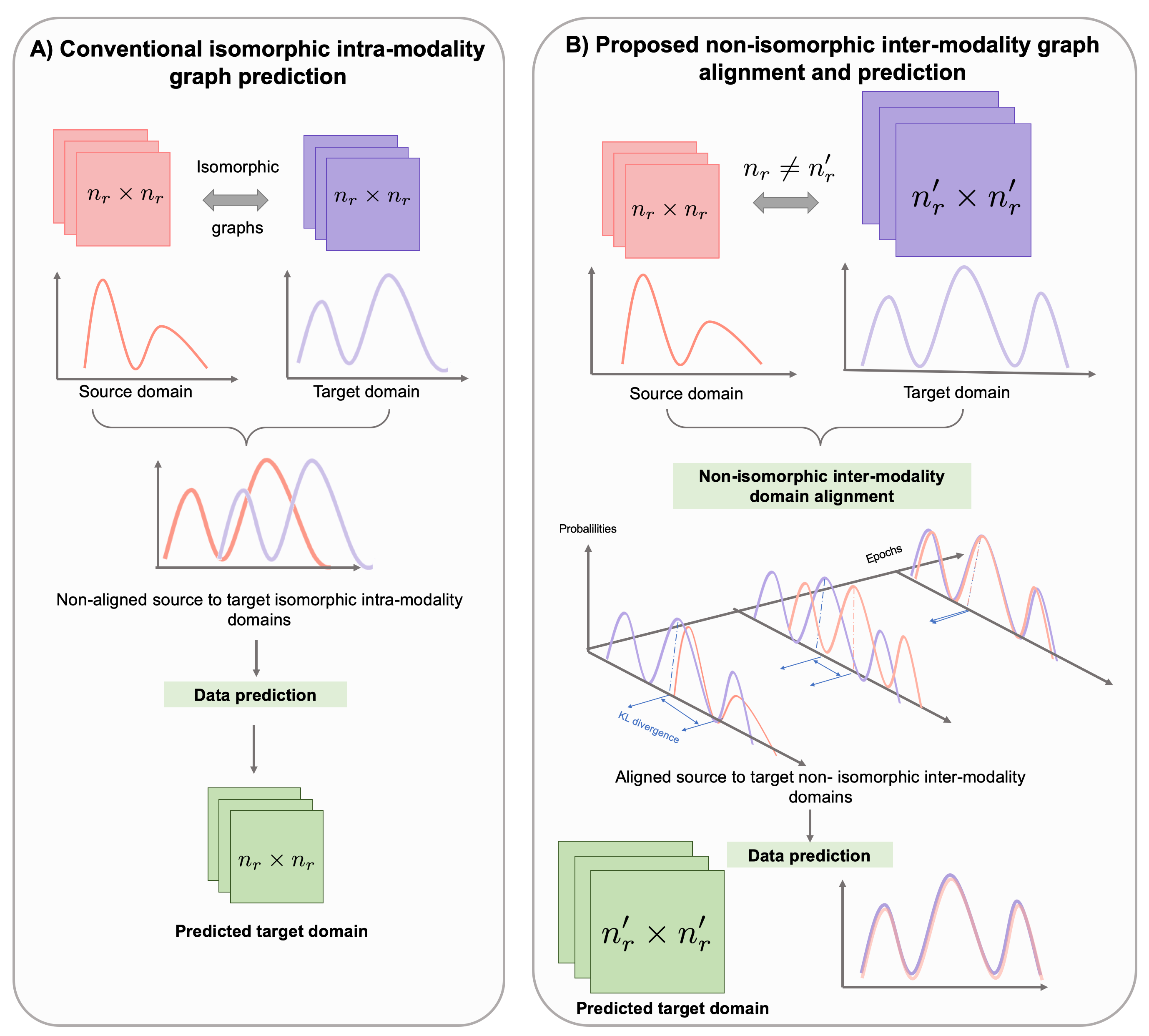}
\caption{\emph{Conventional isomorphic intra-modality graph prediction versus proposed non-isomorphic inter-modality graph alignment and prediction framework.} \textbf{A)} Conventional brain graph synthesis works focus on predicting \emph{isomorphic intra-modality} target graphs without alignment. \textbf{B)} To overcome the limitations of such models, we design a simple but effective \emph{non-isomorphic inter-modality} graph alignment and prediction framework with the following contributions. \emph{First,} we propose a KL divergence-based graph aligner to align the distribution of the training source graphs (from a source modality) to that of the target graphs (from a target modality). \emph{Second}, we design a graph GAN to synthesize a target modality graph from a source one while handling shifts in graph resolution (i.e., node size). \emph{Third,} we design a new Ground Truth-Preserving (GT-P) loss function to guide the \emph{non-isomorphic} generator in learning the topological structure of ground truth target brain graphs more effectively.} 
\label{fig:1}
\end{figure}

Recently, deep learning techniques have achieved great success on graph-structured data which provides a new way to model
the non-linear cross-modality relationship. Specifically, deep graph convolutional networks (GCNs) have permeated the field of brain graph research \cite{isallari2020,nebli2020,bessadok2021} via diverse tasks such as learning the mapping between human connectome and disease diagnosis. Recently, landmark studies used GCNs to predict a target brain graph from a source brain graph. For instance, \cite{zhang2020} proposed a novel GCN model for multimodal brain networks analysis to generate a functional connectome from a structural connectome. Moreover, \cite{zhang2020b} presented a multi-GCN based generative adversarial network (MGCN-GAN) to infer individual structural connectome from a functional connectome. Another recent work \cite{bessadok2020} introduced MultiGraphGAN architecture, which predicts multiple brain graphs from a single brain graph while preserving the topological structure of each target predicted graph. However, all these works use \emph{isomorphic} graphs which means that the source graphs and the ground-truth graphs have the same number of edges, nodes, and are topologically identical. Therefore, using \emph{non-isomorphic} graphs remains a significant challenge in designing generalizable and scalable brain graph synthesis. Moreover, inferring a target domain from a source domain introduces the problem of \emph{domain fracture} resulting in the difference in distribution between the source and target domains. Remarkably, domain alignment is strikingly lacking in brain graph synthesis tasks \cite{zhang2020,zhang2020b,bessadok2020} (\textbf{Fig.}~\ref{fig:1}).
 
To address such unsolved challenges and motivated by the recent development of geometric deep neural network-based solutions, we propose an inter-modality aligner of non-isomorphic brain graphs (IMANGraphNet) framework based on generative adversarial learning. To do so, prior to the prediction block, we propose a graph aligner network to align the training graphs of the source modality to that of the target one. Second, given the aligned source graph, we design a non-isomorphic graph GAN (gGAN) to map the \emph{aligned} source graph from one modality (e.g., morphological) to the target graph modality (e.g., functional). Note that the alignment step facilitates the training of our non-isomorphic gGAN since both source and target domains have been aligned by the aligner network (i.e., shared mode) (\textbf{Fig.}~\ref{fig:2}--A). Besides, in order to capture the complex relationship in both direct and indirect brain connections \cite{zhang2020b}, we design the generator and discriminator of our GAN using GCNs. Moreover, to resolve the inherent instability of GAN, we propose a novel ground-truth-preserving (GT-P) loss function to force our \emph{non-isomorphic} generator to learn the ground-truth brain graph more efficiently. More importantly, by comparing the strongest connectivities of the ground-truth graphs with those of the predicted brain graphs for the same subjects, we investigate the \emph{reproducible} power of our IMANGraphNet which not only predicts reliable brain graphs but also captures the delicate difference across subjects. The compelling aspects of our method can be summarized as follows:
\begin{enumerate}
\item \emph{On a methodological level.} IMANGraphNet presents the first work on inter-modality non-isomorphic graph alignment and synthesis, which can be also leveraged for developing precision medicine using network neuroscience \cite{bassett2017}.

\item \emph{On a clinical level.} Learning brain connectivity inter-modality synthesis can provide holistic brain maps that capture multimodal interactions (functional, structural, and morphological) between brain regions, thereby charting brain dysconnectivity patterns in disordered populations \cite{van2019}. 

\item \emph{On a generic level.} Our framework is a generic method as it can be applied to predict brain graphs derived from any neuroimaging modalities with complex nonlinear distributions even when they are non-isomorphic. IMANGraphNet can also be applied to other types of non-biological graphs.
\end{enumerate}

\section{Methodology}

\begin{figure}[t!]
\centering
\includegraphics[width=12cm]{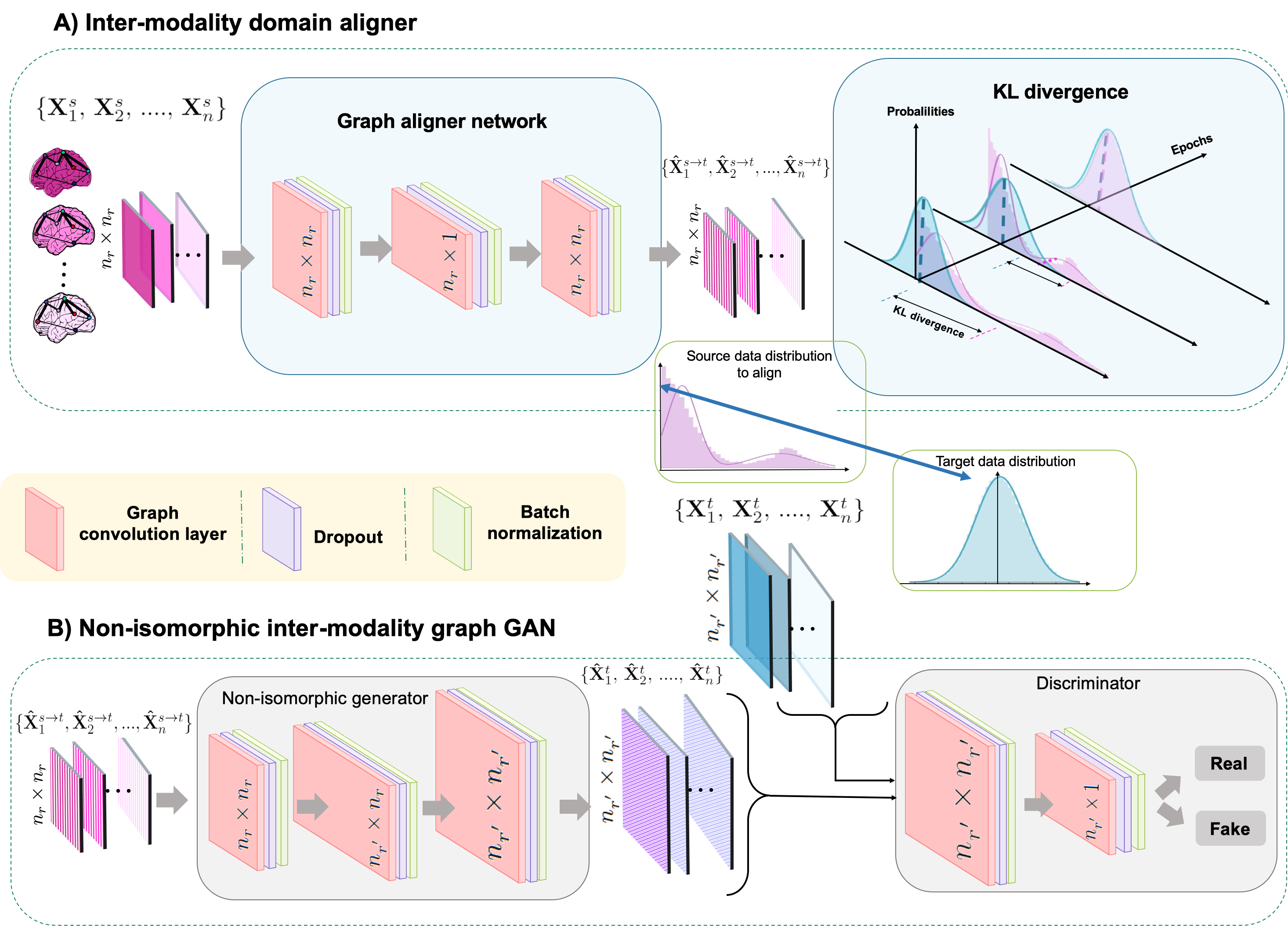}
\caption{ \emph{Illustration of the proposed non-isomorphic inter-modality brain graph alignment and synthesis using IMANGraphNet.} \textbf{A) Graph aligner for inter-modality domain alignment.} We aim to align the training graphs of the source modality $\mathbf{X}^{s}$ to that of the target one $\mathbf{X}^{t}$. Therefore, we design a KL divergence-based graph aligner to bridge the gap between the distributions of the source and target graphs. \textbf{B) Non-isomorphic inter-modality graph GAN.} Next, we propose a non-isomorphic graph GAN to transform the aligned source brain graph $\mathbf{\hat{X}}^{s \rightarrow t}$ (e.g., morphological) into the target graph (e.g., functional) with different structural and topological properties. Both aligner and generator networks are trained in an end-to-end manner by optimizing a novel Ground Truth-Preserving (GT-P) loss function which guides the non-isomorphic generator in learning the topology of the target ground truth brain graphs more effectively.} 
\label{fig:2}
\end{figure}

In this section, we detail our non-isomorphic inter-modality graph alignment and prediction framework (\textbf{Fig.}~\ref{fig:2}). In the first stage, we propose a Kullback-Leibler (KL) divergence-based graph aligner which maps the distribution of the ground truth domain to the source domain. In the second stage, we design a non-isomorphic gGAN to synthesize one modality graph from another while handling graph resolution shifts. Moreover, to handle the unstable behavior of gGAN, we propose a new ground truth-preserving (GT-P) loss function to guide the non-isomorphic generator in learning the topological structure of ground truth brain graphs more effectively. 

$\bullet$ \textbf{Problem statement.} Let $\mathbf{G}_i (\mathbf{V}_i, \mathbf{E}_i)$ denote a brain graph where each node in $\mathbf{V}_i$ denotes a brain region of interest (ROI) and each edge in $\mathbf{E}_i$ connecting two ROIs $k$ and $l$ denotes the strength of their connectivity. Each training subject $i$ in our dataset is represented by two brain graphs $\{\mathbf{G}_{s_i} (\mathbf{V}_{s_i}, \mathbf{E}_{s_i}), \mathbf{G}_{t_i} (\mathbf{V}_{t_i}, \mathbf{E}_{t_i})\}$, where $\mathbf{G}_{s}$ represents the source brain graph with $n_r$ nodes and $\mathbf{G}_{t}$ is the target brain graph with $n_{r^{'}}$ nodes with $n_r \ne n_{r^{'}}$. Specifically, these two graphs are considered as non-isomorphic with no correspondence between nodes and edges across source and target graphs --i.e., they are topologically different (\textbf{Fig.}~\ref{fig:1}). Formally, graph isomorphism can be defined as follows. 

\textbf{Definition 1.} \emph{Two graphs $\mathbf{G}$ and $\mathbf{H}$ are isomorphic if there is a bijection$f : V(\mathbf{G}) \rightarrow V(\mathbf{H})$ so that, for any $v, w \in V(\mathbf{G})$, the number of edges connecting $v$ to $w$ is the same as the number of edges $f(v)$ to $f(w)$. The function $f$ is called an isomorphism from $\mathbf{G}$ to $\mathbf{H}$.}

\textbf{Definition 2.} \emph{Two graphs $\mathbf{G}$ and $\mathbf{H}$ are \emph{non-isomorphic} if they do not satisfy at least one of the following conditions: (i) equal number of nodes, (ii) equal number of edges, and (iii) topologically identical (i.e., preservation of the local neighborhood of each node).}

$\bullet$ \textbf{Graph-based inter-modality aligner block.}
The first block of IMANGraphNet (\textbf{Fig.}~\ref{fig:2}-A) comprises a graph-based inter-modality aligner that constrains the distribution of the mapped source brain graphs to match that of the ground-truth target brain graphs. Inspired by the dynamic edge convolution proposed in \cite{simonovsky2017} and the U-net architecture \cite{ronneberger2015} with skip connections, we propose an aligner network which is composed of three-layer graph convolutional neural network (GCN) (\textbf{Fig.}~\ref{fig:2}-A). Given a set of $n$ training source brain networks (e.g., morphological connectomes) $\mathbf{X}^s_{tr}$ and a set of $n$ training ground-truth brain networks (e.g., functional connectomes) $\mathbf{X}^t_{tr}$, for each subject $i$, our aligner takes $\mathbf{X}^s_i$ as input and outputs $\mathbf{\hat{X}}_i^{s \rightarrow t}$ which shares the same distribution of $\mathbf{X}^t_i$. Our model consists of three GCN layers adjusted by adding batch normalization and dropout to the output of each layer. Specifically, batch normalization efficiently accelerates the network training through a fast convergence of the loss function and dropout eliminates the risk of overfitting. Hence, these two operations help optimize and simplify the network training.

To improve the quality of the inter-modality aligner network, we propose to minimize the discrepancy between ground-truth and aligned source brain graph distributions using KL divergence as a loss function. In fact, KL divergence, also known as the relative entropy, is an asymmetric measure that quantifies the difference between two probability distributions. Thereby, we define our inter-modality graph alignment loss function using KL divergence to minimize the gap between the distributions of the aligned source graphs and that of ground-truth target graphs. Specifically, we compute the KL divergence between the ground truth distribution $q_{tr}$ and aligned distribution $p_{tr}$ for the training subjects which is expressed as follows:

\begin{equation}\mathcal{L}_{KL}= \sum_{i=1}^{n} KL\left(q_{i} \| p_{i}\right)\end{equation}

where the KL divergence for subject $i$ is defined as:
$ KL\left(q_{i} \| p_{i}\right)=\int_{-\infty}^{+\infty} q_{i}(x) \log \frac{q_{i}(x)}{p_{i}(x)} d x$

Note that the KL divergence $KL\left(q_{i} \| p_{i}\right)$ is not a symmetrical function $KL\left(q_{i} \| p_{i}\right) \ne KL\left(p_{i} \| q_{i}\right)$ and defines the information gained by changing beliefs from a prior probability distribution $p$ to the posterior probability distribution $q$ (i.e., moving the prior distribution towards the posterior one). Intuitively, $q$ is the true distribution and $p$ is the aligned.

$\bullet$ \textbf{Adversarial non-isomorphic graph generator block.}
Following the alignment step, we design a \emph{non-isomorphic} generator architecture that handles shifts in graph resolution (i.e., node size variation) coupled with an adversarial discriminator.

\emph{\textbf{Non-isomorphic brain graph generator.}} Our non-isomorphic graph generator $G$ is composed of three GCN layers regularized using batch normalization and dropout to the output of each layer (\textbf{Fig.}~\ref{fig:2}-B), taking as input the aligned source graphs to the target distribution $\mathbf{\hat{X}}^{s \rightarrow t}_i$ of size $n_r \times n_r$ and outputting the predicted target brain graphs $\mathbf{\hat{X}}^t_{i}$ of size $n_{r^{'}} \times n_{r^{'}}$ where $n_{r} \ne n_{r^{'}}$. Specifically, owing to dynamic graph-based edge convolution operation \cite{simonovsky2017}, each GCN layer includes a unique dynamic filter that outputs edge-specific weight matrix which dictates the information flow between nodes $k$ and $l$ to learn a comprehensive vector representation for each node. Next, to learn our inter-modality non-isomorphic mapping, we define a mapping function $\mathcal{T}_{r}: \mathbb{R}^{n_r \times n_{r^{'}}} \mapsto \mathbb{R}^{n_{r^{'}} \times n_{r^{'}}}$ that takes as input the embedded matrix of the whole graph in the latest GCN layer of size ${n_r \times n_{r^{'}}}$ and outputs the generated target graph of size $n_{r^{'}} \times n_{r^{'}}$ (see subsection: \emph{Graph resolution shift based on dynamic edge convolution}).

\emph{\textbf{Graph discriminator based on adversarial training.}} Our non-isomorphic generator $G$ is trained in an adversarial manner against a discriminator network $D$ (\textbf{Fig.}~\ref{fig:2}-B). In order to discriminate between the predicted and ground truth target graph data, we design a two-layer graph neural network \cite{simonovsky2017}. Our discriminator $D$ takes as input the real connectome $\mathbf{X}^t_i$ and the generator's output $\mathbf{\hat{X}}^t_i$. The discriminator outputs a value between $0$ and $1$ measuring the realness of the generator's output. To enhance our discriminator's ability to distinguish between the target predicted and ground truth brain graphs, we adopt the adversarial loss function so that it maximizes the discriminator's output value for the $\mathbf{X}^t_i$ and minimizes it for $\mathbf{\hat{X}}^t_i$.

\textbf{\emph{Graph resolution shift based on dynamic edge convolution.}} In all network blocks of IMANGraphNet, each proposed GCN layer uses a dynamic graph-based edge convolution process \cite{simonovsky2017}. Specifically, let $h$ be the layer index in the neural network and $d_h$ denote the output dimension of the corresponding layer. Each layer $h$ includes a filter generating network $F^{h} : \mathbb{R} \mapsto \mathbb{R}^{d_{h} \times d_{h - 1}}$ that dynamically generates a weight matrix for filtering message passing between ROIs $k$ and $l$ given the edge weight $e_{kl}$. Here $e_{kl}$ is edge feature (i.e., connectivity weight) that quantifies the relationship between ROIs $k$ and $l$. The purpose of each layer in our IMANGraphNet is to produce the graph convolution result which can be observed as a filtered signal $\mathbf{z}^{h}(k) \in \mathbb{R}^{d_{h}\times 1}$ at node $k$. The overall edge-conditioned convolution operation is defined as follows:
\begin{equation}
\mathbf{z}{_{k}^{h}} = \mathbf{\Theta}^{h} .\mathbf{z}{_{k}^{h-1}} + \frac{1}{\left |N(k)\right |}\sum_{l \epsilon N(k)} \ F^{h}(e_{kl}; \mathbf{W}^{h}) \mathbf{z}^{h - 1}_{l} + \mathbf{b}^{h}
\end{equation}
where $\mathbf{z}{_{k}^{h}} \in \mathbb{R}^{d_h \times 1}$ is the embedding of node $k$ in layer $h$, $\Theta_{l k}^h=F^{h}\left(\mathbf{e}_{kl}; \mathbf{W}^{h}\right)$ represents the dynamically generated edge-specific weights by $F^h$. $\mathbf{b}^{h} \in \mathbb{R}^{d_{h}}$ denotes a network bias and $N(k)$ denotes the neighbors of node $k$.

Given the learned embedding $\mathbf{z}{_{k}^{h}} \in \mathbb{R}^{d_h }$ for node $k$ in layer $h$, we define the embedding of the whole graph in layer $h$ as $\mathbf{Z}^{h} \in \mathbb{R}^{n_r \times d_h}$ where $n_r$ is the number of nodes. We draw to the attention of the reader that any resolution shift can be easily expressed as a transformation $\mathcal{T}_{r} : \mathbb{R}^{n_r \times d_h} \mapsto \mathbb{R}^{d_{h} \times d_{h}}$ where $\mathcal{T}_{r}$ is formulated as follows: $\mathcal{T}_{r} = ({\mathbf{Z}^{h}})^T \mathbf{Z}^{h}$. As such, shifting resolution is only defined by fixing the desired target graph resolution $d_{h}$. In our case, we set $d_h$ of the latest layer in the generator to $n_{r^{'}}$ to output the predicted target brain graph $\mathbf{\hat{X}}^t$ of size $n_{r^{'}} \times n_{r^{'}}$ (\textbf{Fig.}~\ref{fig:2}).

\emph{\textbf{Ground truth-Preserving loss function.}}
GAN generators are conventionally optimized according to the response of their corresponding discriminators. However, within a few training epochs, we note that the discriminator can easily distinguish real graphs from predicted graphs and the adversarial loss would be close to 0. In this case, the generator cannot provide good results and will keep producing bad quality graphs. To overcome this issue, we need to enforce a synchrony between the generator and the discriminator learning throughout the whole training process.
Thus, we propose a new ground truth-preserving (GT-P) loss function composed of four sub-losses: adversarial loss, $L1$ loss, Pearson correlation coefficient (PCC) loss and topological loss, which we detail below. We define our GT-P loss function as follows: 
\begin{equation}
\mathcal{L}_{\text {GT-P}}= \lambda_1 \mathcal{L}_{adv}+\lambda_2 \mathcal{L}_{L 1}+\lambda_3 \mathcal{L}_{PCC}+\lambda_4 \mathcal{L}_{top}
\end{equation}

where $\mathcal{L}_{a d v}$ represents the adversarial loss which quantifies the difference between the generated and ground truth target graphs as both non-isomorphic generator and discriminator are iteratively optimized through the adversarial loss:

\begin{equation}
\operatorname{argmin}_{G} \operatorname{\max}_{D} \mathcal{L}_{a d v}=\mathbb{E}_{G\left(\mathbf{X}^t\right)}\left[\log\left(D\left(G\left(\mathbf{X}^t\right)\right)\right]+\mathbb{E}_{G\left(\mathbf{\hat{X}}^t\right)}\left[\log \left(1-D\left(G\left(\mathbf{\hat{X}}^t\right)\right)\right]\right.\right.
\end{equation}

To improve the quality of the predicted target brain graphs, we propose to add an $l1$ loss term that minimizes the distance between each predicted subject $\mathbf{\hat{X}}^{t}$ and its related ground truth $\mathbf{X}^{t}$. The $l1$ loss function is expressed as follows:
$\mathcal{L}_{l 1}=\left\|{\mathbf{X}}^{t} - \mathbf{\hat{X}}^{t} \right\|_{1}.$

Even robust to outliers, the $l1$ loss only focuses on the element-wise similarity in edge weights between the predicted and real brain graphs and ignores the overall correlation between both graphs. Hence, we include the Pearson correlation coefficient (PCC) in our loss which measures the overall correlation between the predicted and real brain graphs. Since (i) the non-isomorphic generator aims to minimize its loss function and (ii) higher PCC indicates a higher correlation between the ground-truth and the predicted graphs, we propose to minimize the PCC loss function as follows: $\mathcal{L}_{PCC}= 1 - PCC$.

We further note that each brain graph has its unique topology which should be preserved when generating the target brain graphs. Therefore, we introduce a topological loss function that forces the non-isomorphic generator to maintain the nodes' topological profiles while learning the global graph structure. To do so, we first compute eigenvector centrality (capturing the centralities of a node's neighbors) of each node for both predicted and real brain graphs. Then, we define the $l1$ loss between the real and predicted eigenvector centralities in order to minimize the discrepancy between them. Hence, we define our topology loss as $\mathcal{L}_{top}=\left\|\mathbf{c}^{t}-\mathbf{\hat{c}}^{t}\right\|_{1}$, where $\mathbf{\hat{c}}^{t}$ denotes the eigenvector centrality vector of the predicted brain graph and $\mathbf{c}^{t}$ is the eigenvector centrality vector of the real one.

\section{Experimental results and discussion}

\textbf{Evaluation dataset.} We used three-fold cross-validation to evaluate the proposed IMANGrahNet framework on 150 subjects from the Southwest University Longitudinal Imaging Multimodal (SLIM) public dataset\footnote{\url{http://fcon\_1000.projects.nitrc.org/}} where each subject has T1-w, T2-w MRI and resting-state fMRI (rsfMRI) scans. Our IMANGraphNet is implemented using PyTorch-Geometric library \cite{fey2019}.

\emph{Morphological brain networks (source).} We used FreeSurfer \cite{fischl2012} to reconstruct the cortical morphological network for each subject from structural T1-w MRI. Specificlaly, we parcellated each cortical hemisphere into 35 cortical regions using Desikan-Killiany cortical atlas. Finally, by computing the pairwise absolute difference in cortical thickness between pairs of regions of interest, we generated a $35 \times 35$ morphological connectivity matrix for each subject denoted as $\mathbf{X}^s$.

\emph{Functional brain networks (target).} Following several preprocessing steps of each resting-state fMRI using preprocessed connectomes project quality assessment protocol, brain graphs (connectomes) were produced using a whole-brain parcellation approach as proposed in \cite{dosenbach2010}. Each brain rfMRI was partitioned into 160 ROIs. Functional connectivity weights were computed using the Pearson correlation coefficient between two average fMRI signals of pairs of ROIs. These denote our target brain graphs $\mathbf{X}^t$.

\textbf{Parameter setting.} For the hyperparameters of the aligner network, we set $\lambda_{KL} = 0.001$. Also, we set the non-isomorphic generator's hyperparameters as follows: $\lambda_1 = 1$, $\lambda_2 = 1$, $\lambda_3 = 0.1$, and $\lambda_4 = 2$. Moreover, we chose AdamW \cite{loshchilov2018} as our default optimizer and set the learning rate at $0.025$ for both the aligner and the non-isomorphic generator networks and $0.01$ for the discriminator. Finally, we trained our IMANGraphNet for 400 epochs using a single Tesla V100 GPU (NVIDIA GeForce GTX TITAN with 32 GB memory). The feature dimensions of GCNs in the aligner are: GCN1 = (35, 35), GCN2 = (35, 1), GCN3 = (35, 35). The feature dimensions of GCNs in the non-isomorphic generator are: GCN1 = (35, 35), GCN2 = (35, 160), GCN3 = (160, 160). Similarly, feature dimensions of GCNs in the discriminator are set to GCN1 = (160, 160) and GCN2 = (160, 1).

\begin{figure}[ht]
\centering
\includegraphics[width=12cm]{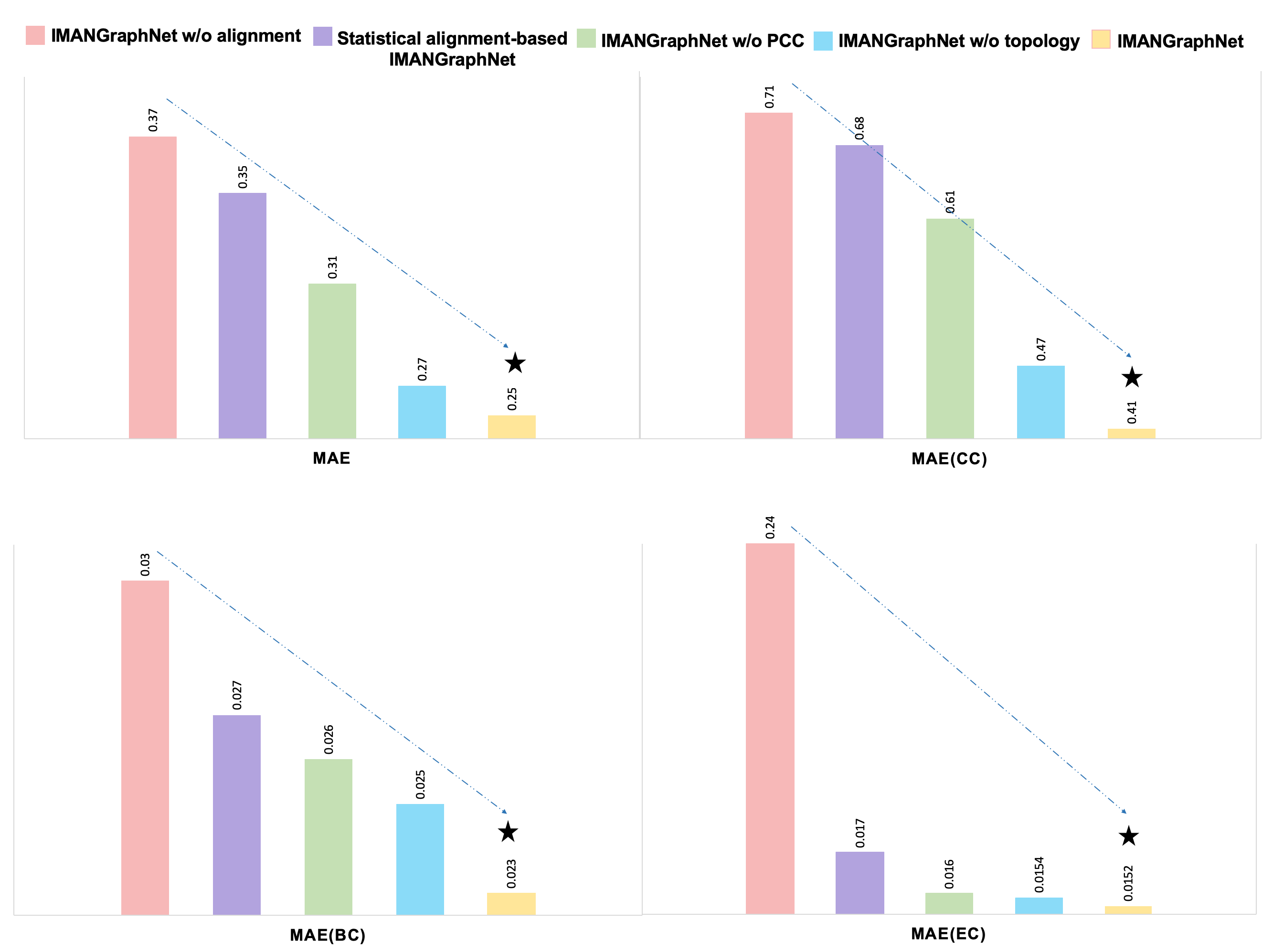}
\caption{\emph{Prediction results using different evaluation metrics.} Evaluation of alignment and prediction brain graph synthesis by our framework IMANGraphNet against four comparison methods: (1) IMANGraphNet w/o alignment, (2) Statistical alignment-based IMANGraphNet, (3) IMANGraphNet w/o PCC and (4) IMANGraphNet w/o topology. As evaluation metrics, we used the mean absolute error (MAE) between the target ground truth and predicted brain graphs as well as their mean absolute difference in three topological measures (CC: closeness centrality, BC: betweenness centrality and EC: eigenvector centrality). w/o: without. $\star$: Our method IMANGrapNet \emph{significantly} outperformed all benchmark methods using two-tailed paired t-test ($p < 0.05$) --excluding the statistical alignment-based IMANGraphNet using MAE(EC).} 
\label{fig:3}
\end{figure}

\begin{figure}[ht!]
\centering
\includegraphics[width=12cm]{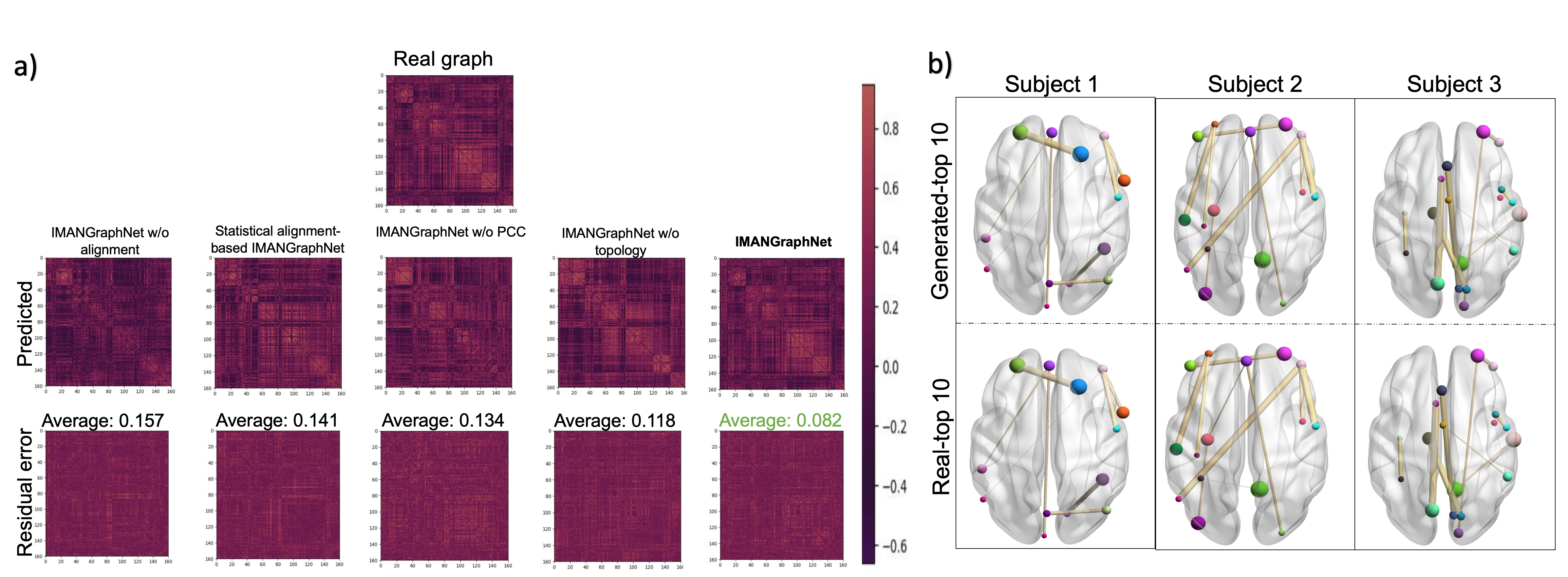}
\caption{\emph{Visual comparison between the real and the predicted target brain graphs.} a) Comparison between the ground truth and predicted brain graphs by IMANGraphNet and
four baseline methods (IMANGraphNet w/o alignment, Statistical alignment-basedIMANGraphNet, IMANGraphNet w/o PCC, IMANGraphNet w/o topology) using a representative testing subject. We display the residual matrices computed using the absolute difference between ground truth and predicted brain graph connectivity matrices. b) The top 15 strongest connectivities of real and predicted functional brain networks of 3 randomly selected testing subjects.} 
\label{fig:4}
\end{figure}

\textbf{Evaluation and comparison methods.} To evaluate the effectiveness of our proposed method for predicting one modality from another, we carried out four major comparisons:
(1) IMANGraphNet w\textbackslash o alignment which considers only the prediction task without aligning the distribution of the source graph to the ground truth graph. (2) Statistical alignment-based IMANGraphNet where we used a statistical alignment instead of the learned alignment when matching the distribution of the source graphs to the ground truth graphs. (3) IMANGraphNet w\textbackslash o PCC where we used our proposed framework without the Pearson correlation coefficient based loss. (4) IMANGraphNet w\textbackslash o topology where we used our proposed framework without any topological loss. As illustrated in \textbf{Fig.}~\ref{fig:3}, we computed the mean absolute error between the target predicted brain graphs and the real brain graphs. Clearly, our IMANGraphNet \emph{significantly} ($p-value < 0.05$ using two-tailed paired t-test) outperformed comparison methods by achieving the lowest error between the predicted and real brain graphs across all evaluation metrics including topological properties using closeness centrality, betweenness centrality and eigenvector centrality.

\emph{Brain graph alignment.} As it is shown in \textbf{Fig.}~\ref{fig:3}, IMANGraphNet w\textbackslash o alignment method achieved the highest (MAE) between the real and predicted brain graphs. This shows that the domain alignment improves the quality of the generated brain graphs in the target modality domain (i.e., functional). We also notice an improvement in performance when using a simple statistical alignment strategy (statistical alignment-based IMANGraphNet) despite the inherent assumption that both source and target distributions are normal (a bell-shaped curve). Hence, when using a complex non-linear distribution (as our source morphological distribution in \textbf{Fig.}~\ref{fig:2}), the statistical aligner cannot align to the target distribution properly. Undeniably, a \emph{learning-based} aligner has the ability to better adapt to any kind of distribution, thereby achieving the best inter-modality graph synthesis results.

\emph{Insights into topological measures.} To investigate the fidelity of the predicted brain graphs to the real brain graphs in topology and structure, we evaluated our method using different topological measures (eigenvector, closeness, and betweenness). As shown in \textbf{Fig.}~\ref{fig:3}, our framework produced the smallest MAE between the functional ground truth and predicted brain graphs across all topological measurements. This shows that our IMANGraphNet is able to well preserve the most central and important nodes (i.e., hub nodes) in the synthesized functional connectomes.

\emph{Insights into the proposed loss objective.}
To prove the superiority of the proposed GT-P loss function, we trained our IMANGraphNet with different loss functions. As illustrated in \textbf{Fig.}~\ref{fig:3}, our GT-P loss function outperforms its ablated versions. These results can be explained by the fact that the $l1$ loss focuses only on minimizing the distance between two brain graphs at the local level. Besides, PCC captures the strength of the correlation between both brain graphs. It aims to maximize the similarity of global connectivity patterns between the predicted and real brain graphs. However, both losses overlook the topological properties of brain graphs (e.g., node centrality). For this reason, we introduced the eigenvector centrality in our topological loss which quantifies the influence of a node on information flow in a network. The combination of these complementary losses achieved the best functional connectivity prediction results from morphological connectivity while relaxing the graph isomorphism (\textbf{Definition 1}) assumption between source and target domains. 

\emph{Reproducibility.} In addition to generating realistic functional brain graphs, our framework could also capture the delicate differences in connectivity patterns across subjects. Specifically, we display in \textbf{Fig.}~\ref{fig:4}-a the real, predicted, and residual brain graphs for a representative testing subject using five different methods. The residual graph is calculated by taking the absolute difference between the real and predicted brain graphs. An average difference value of the residual is displayed on top of each residual graph. We observe that the residual was noticeably reduced by our IMANGraphNet method.

\emph{Clinical interest.} \textbf{Fig.}~\ref{fig:4}-b displays the top $10$ strongest connectivities of real and predicted functional brain graphs of $3$ randomly selected testing subjects. Since brain connectivity patterns vary across different individuals \cite{glasser2016}, we notice that the top $10$ connectivities are not identical. However, our model can reliably predict such variations as well as individual trends in functional connectivity based solely on morphological brain graphs derived from T1-w MRI. This result further confirms that our approach is trustworthy for predicting \emph{multimodal} brain dysconnectivity patterns in disordered populations \cite{van2019} from limited neuroimaging resources.

\section{Conclusion}
In this paper, we introduced the first geometric deep learning architecture, namely IMANGraphNet, for \emph{inter-modality non-isomorphic} brain graph synthesis, which nicely handles variations in graph distribution, size and structure. Our key contributions consist in designing: (1) a graph aligner network to align the training graphs of the source modality to that of the target one and (2) a non-isomorphic generator architecture that handles shifts in graph resolution (i.e., node size variation) coupled with an adversarial discriminator using GCNs.
Furthermore, we proposed a new ground truth-preserving loss function which guides the non-isomorphic generator in learning the topology of the target ground truth brain graphs more effectively. Our framework outperforms the baseline methods in terms of alignment and prediction results. IMANGraphNet not only predicts reliable functional brain graphs from morphological ones but also preserves the topology of the target domain. In our future work, we will extend our architecture to predict \emph{multiple} modality graphs from a single source one.

\section{Acknowledgments}

This work was funded by generous grants from the European H2020 Marie Sklodowska-Curie action (grant no. 101003403, \url{http://basira-lab.com/normnets/}) to I.R. and the Scientific and Technological Research Council of Turkey to I.R. under the TUBITAK 2232 Fellowship for Outstanding Researchers (no. 118C288, \url{http://basira-lab.com/reprime/}). However, all scientific contributions made in this project are owned and approved solely by the authors.

\bibliography{Biblio3}

\begin{thebibliography}{10}

\bibitem{shen2013}
Shen, L., Liu, T., Yap, P.T., Huang, H., Shen, D., Westin, C.F.:
\newblock Multimodal brain image analysis: Third international workshop, mbia
  2013, held in conjunction with miccai 2013, nagoya, japan, september 22,
  2013, proceedings.
\newblock \textbf{8159} (2013)

\bibitem{zhang2019}
Zhang, Y., Huang, H.:
\newblock New graph-blind convolutional network for brain connectome data
  analysis.
\newblock International Conference on Information Processing in Medical Imaging
  (2019)  669--681

\bibitem{yu2020}
Yu, B., Wang, Y., Wang, L., Shen, D., Zhou, L.:
\newblock Medical image synthesis via deep learning.
\newblock Deep Learning in Medical Image Analysis (2020)  23--44

\bibitem{zhou2020}
Zhou, T., Fu, H., Chen, G., Shen, J., Shao, L.:
\newblock Hi-net: hybrid-fusion network for multi-modal mr image synthesis.
\newblock IEEE transactions on medical imaging (2020)

\bibitem{liu2020}
Liu, Y., Pan, Y., Yang, W., Ning, Z., Yue, L., Liu, M., Shen, D.:
\newblock Joint neuroimage synthesis and representation learning for conversion
  prediction of subjective cognitive decline.
\newblock International Conference on Medical Image Computing and
  Computer-Assisted Intervention (2020)  583--592

\bibitem{dai2020}
Dai, X., Lei, Y., Fu, Y., Curran, W.J., Liu, T., Mao, H., Yang, X.:
\newblock Multimodal mri synthesis using unified generative adversarial
  networks.
\newblock Medical Physics (2020)

\bibitem{yang2020}
Yang, Q., Li, N., Zhao, Z., Fan, X., Eric, I., Chang, C., Xu, Y.:
\newblock Mri cross-modality image-to-image translation.
\newblock Scientific Reports \textbf{10} (2020)  1--18

\bibitem{bassett2017}
Bassett, D.S., Sporns, O.:
\newblock Network neuroscience.
\newblock Nature neuroscience \textbf{20} (2017)  353--364

\bibitem{van2019}
van~den Heuvel, M.P., Sporns, O.:
\newblock A cross-disorder connectome landscape of brain dysconnectivity.
\newblock Nature reviews neuroscience \textbf{20} (2019)  435--446

\bibitem{bronstein2017}
Bronstein, M.M., Bruna, J., LeCun, Y., Szlam, A., Vandergheynst, P.:
\newblock Geometric deep learning: going beyond euclidean data.
\newblock IEEE Signal Processing Magazine \textbf{34} (2017)  18--42

\bibitem{isallari2020}
Isallari, M., Rekik, I.:
\newblock Gsr-net: Graph super-resolution network for predicting
  high-resolution from low-resolution functional brain connectomes.
\newblock International Workshop on Machine Learning in Medical Imaging (2020)
  139--149

\bibitem{nebli2020}
Nebli, A., Kaplan, U.A., Rekik, I.:
\newblock Deep evographnet architecture for time-dependent brain graph data
  synthesis from a single timepoint.
\newblock International Workshop on PRedictive Intelligence In MEdicine (2020)
  144--155

\bibitem{bessadok2021}
Bessadok, A., Mahjoub, M.A., Rekik, I.:
\newblock Brain graph synthesis by dual adversarial domain alignment and target
  graph prediction from a source graph.
\newblock Medical Image Analysis \textbf{68} (2021)  101902

\bibitem{zhang2020}
Zhang, W., Zhan, L., Thompson, P., Wang, Y.:
\newblock Deep representation learning for multimodal brain networks.
\newblock International Conference on Medical Image Computing and
  Computer-Assisted Intervention (2020)  613--624

\bibitem{zhang2020b}
Zhang, L., Wang, L., Zhu, D.:
\newblock Recovering brain structural connectivity from functional connectivity
  via multi-gcn based generative adversarial network.
\newblock International Conference on Medical Image Computing and
  Computer-Assisted Intervention (2020)  53--61

\bibitem{bessadok2020}
Bessadok, A., Mahjoub, M.A., Rekik, I.:
\newblock Topology-aware generative adversarial network for joint prediction of
  multiple brain graphs from a single brain graph.
\newblock International Conference on Medical Image Computing and
  Computer-Assisted Intervention (2020)  551--561

\bibitem{simonovsky2017}
Simonovsky, M., Komodakis, N.:
\newblock Dynamic edge-conditioned filters in convolutional neural networks on
  graphs.
\newblock Proceedings of the IEEE conference on computer vision and pattern
  recognition (2017)  3693--3702

\bibitem{ronneberger2015}
Ronneberger, O., Fischer, P., Brox, T.:
\newblock U-net: Convolutional networks for biomedical image segmentation.
\newblock International Conference on Medical image computing and
  computer-assisted intervention (2015)  234--241

\bibitem{fey2019}
Fey, M., Lenssen, J.E.:
\newblock Fast graph representation learning with pytorch geometric.
\newblock arXiv preprint arXiv:1903.02428 (2019)

\bibitem{fischl2012}
Fischl, B.:
\newblock Freesurfer.
\newblock Neuroimage \textbf{62} (2012)  774--781

\bibitem{dosenbach2010}
Dosenbach, N.U., Nardos, B., Cohen, A.L., Fair, D.A., Power, J.D., Church,
  J.A., Nelson, S.M., Wig, G.S., Vogel, A.C., Lessov-Schlaggar, C.N.,  et~al.:
\newblock Prediction of individual brain maturity using fmri.
\newblock Science \textbf{329} (2010)  1358--1361

\bibitem{loshchilov2018}
Loshchilov, I., Hutter, F.:
\newblock Fixing weight decay regularization in adam.
\newblock (2018)

\bibitem{glasser2016}
Glasser, M., Coalson, T., Robinson, E., Hacker, C.D., \emph{et al.}:
\newblock A multi-modal parcellation of human cerebral cortex.
\newblock Nature \textbf{536} (2016)  171--178

\end{thebibliography}
\bibliographystyle{splncs}
\end{document}